\documentclass[12pt]{article}

\pdfoutput=1

\usepackage{amsmath,amssymb}
\usepackage{bm}
\usepackage{graphicx}

\makeatletter

\@addtoreset{equation}{section}
\makeatother

\newcommand{\beq}{\begin{equation}}
\newcommand{\eeq}{\end{equation}}

\newcommand{\bR}{\ensuremath{\mathbb{R}}}

\newcommand{\bZ}{\ensuremath{\mathbb{Z}}}

\begin{document}

\baselineskip=18pt  
\baselineskip 0.8cm

\begin{titlepage}

\setcounter{page}{0}

\renewcommand{\thefootnote}{\fnsymbol{footnote}}

\begin{flushright}
{\tt CALT-68-2754\\
IPMU09-0129\\
KUNS-2236}
\end{flushright}

\vskip 1cm

\begin{center}
{\LARGE \bf
Gravity Dual of Spatially Modulated Phase}

\vskip 2cm

{
{\large Shin Nakamura$^1$, Hirosi Ooguri$^{2,3}$ and Chang-Soon Park$^2$}
}

\vskip 1cm

{
\it
$^1$Department of Physics, Kyoto University,
Kyoto 606-8502, Japan\\
\vskip 0.05cm
$^2$California Institute of Technology, Pasadena, CA 91125, USA\\
\vskip 0.05cm
$^3$ IPMU, University of Tokyo, Kashiwa 277-8586, Japan\\}

\end{center}

\vspace{1.5cm}

\centerline{{\bf Abstract}}
\medskip
\noindent
We show that the five-dimensional Maxwell theory with the Chern-Simons 
term is tachyonic in the presence of a constant electric field.  
When coupled to gravity, a sufficiently large Chern-Simons coupling 
causes instability of the Reissner-Nordstr\"om black holes in anti-de Sitter 
space. The instability happens only at non-vanishing momenta, suggesting 
a spatially modulated phase in the holographically dual quantum field 
theory in ($3+1$) dimensions, with spontaneous current generation in a 
helical configuration. 
The three-charge extremal black hole in the type IIB
superstring theory on $AdS_5 \times S^5$ barely satisfies the stability
condition. 

\end{titlepage}
\setcounter{page}{1} 


\section{Introduction}

Instability of black holes in anti-de Sitter space has attracted much
attention recently due to its relevance to quantum phase transitions in dual
strongly interacting quantum field theory at finite density
 \cite{Hartnoll, Herzog}. In this paper, we point out 
a new type of instability caused by the Chern-Simons term. 
A novel feature is that the instability happens only at non-vanishing
momenta, suggesting a spatially modulated phase transition in
the holographically dual field theory.  

In three dimensions, the Maxwell theory becomes massive when 
the Chern-Simons term is included \cite{Deserone,Desertwo}. In 
higher dimensions, the Chern-Simons term starts with a higher power
in gauge fields, but it can contribute to quadratic fluctuations 
if there is a non-zero background gauge field. 
In this paper, we will show that the Maxwell theory in five dimensions
with the Chern-Simons term becomes tachyonic if we turn on a constant electric 
field. In contrast, a background magnetic field does not cause instability,
but it makes the gauge field massive as in three dimensions.\footnote{To our
knowledge, \cite{Domokos} is the first paper to point out that the 
Chern-Simons term in five dimensions induces instability. They considered
a system consisting of two non-Abelian gauge fields coupled to an adjoint 
scalar field with a tachyonic mass as a holographic model of QCD and reduced it to
four dimensions before studying its spectrum. Though their set-up and 
analysis are 
different from ours, the dispersion relation we derive in section \ref{S:MaxwellCS} is related to theirs.
We will point this out at an appropriate place in section \ref{S:MaxwellCS}.}

Chern-Simons terms abound in supergravity theories, and 
charged black hole solutions in these theories
provide an interesting laboratory in which to study the instability and its 
implications since these solutions carry background electric fields. 
The near-horizon geometry of the five-dimensional 
extremal Reissner-Nordstr\"om black hole in $AdS_5$ is 
$AdS_2 \times \bR^3$ with the gauge field strength 
proportional to the volume form of $AdS_2$. The background electric field 
causes mixing of the gauge field with 
the metric at the quadratic order, and we will take it into account 
in our stability analysis. We find a critical value $\alpha_{{\rm crit}}$ 
of the Chern-Simons coupling $\alpha$ above which the near-horizon 
geometry becomes unstable for some range of momenta $k$ in $\bR^3$.
Interestingly, the range excludes $k=0$, $i.e.$ the instability happens 
only at non-zero spatial momentum.

\begin{figure}[ht]
\centering
\includegraphics[width=7cm]{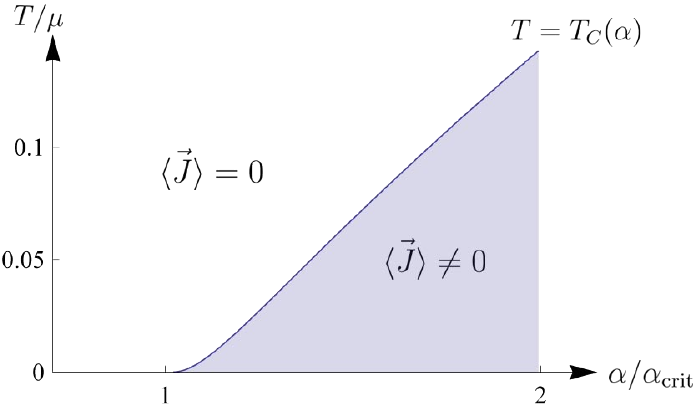}
\caption{Critical temperature as a function of the Chern-Simons coupling $\alpha$. 
The shaded region indicates a phase with a non-zero expectation value of the 
conserved current $\vec{J}$ which is helical and position dependent.}\label{criticaltemperature}
\end{figure}

\begin{figure}[ht]
\begin{center}$
\begin{array}{cc}
\includegraphics[height=5cm]{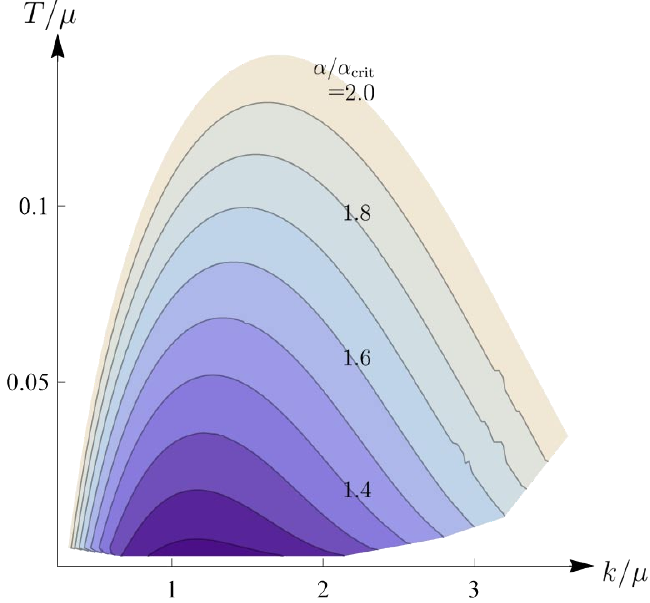} &
\includegraphics[height=5cm]{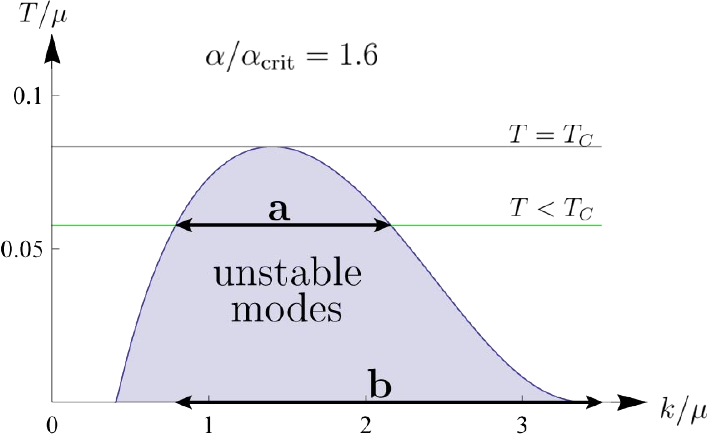}
\end{array}$
\end{center}
\caption{The left figure indicates unstable regions 
for various values of the Chern-Simons coupling $\alpha$. 
The right figure is for a particular choice of the Chern-Simons 
coupling $\alpha=1.6 \alpha_{\rm crit}$. 
The critical temperature $T_C$ is the maximum temperature 
with unstable modes.
The figure indicates the unstable range $\textbf{a}$ 
for some temperature $T < T_C$. 
The range $\textbf{b}$ is derived from the near-horizon 
analysis at $T=0$. Note that the actual range 
of unstable momenta is wider.}
\label{aTkgraph}
\end{figure}

The Reissner-Nordstr\"om black hole solution in $AdS_5$ 
gives a holographic description of a thermodynamic state in the dual
conformal field theory at finite temperature $T$ and chemical potential 
$\mu$.\footnote{\cite{Chamblin:1999tk,Cvetic:1999ne} studied the thermodynamic properties of the Reissner-Nordstr\"om AdS black hole. Its relation to Fermi liquid is discussed in \cite{Rey:2008zz}.} We find that, for $\alpha > \alpha_{{\rm crit}}$, there is a
critical temperature $T_{{\rm c}}(\alpha)$ below which 
the black hole solution becomes unstable, as shown in Fig. \ref{criticaltemperature}.
The instability happens at a range of momenta, which becomes 
wider as $T$ is lowered
but never includes $k=0$, as shown in Fig. \ref{aTkgraph}. 
We find an interesting subtlety 
in the zero temperature limit; the unstable range is 
wider than the range expected from the analysis
near the horizon of the extremal black hole. 
It turns out that the near-horizon analysis gives a sufficient but
not necessary condition since there are unstable modes 
in the full Reissner-Nordstr\"om solution which do not reduce 
to normalizable modes in $AdS_2 \times \bR^3$ in the near-horizon limit.

In the dual field theory in $(3+1)$ dimensions, the instability of the 
Reissner-Nordstr\"om solution can be interpreted as a signal of 
 a novel phase transition at finite 
chemical potential where the charge current $\vec{J}(x)$ dual 
to the gauge field
develops a position dependent expectation value of the form,
\beq
  \langle \vec{J} (x) \rangle = {\rm Re}\left( \vec{u} e^{ikx}\right),
\label{VEV}
\eeq
with non-zero momentum $k$. The constant vector $\vec{u}$ is circularly
polarized as,
\beq 
  \vec{k} \times  \vec{u} = \pm i |k|\vec{u},
\eeq
where the sign is correlated to the sign of the Chern-Simons couping
as we will explain later. The vacuum expectation value \eqref{VEV}
is helical and breaks translational and rotational symmetries
in three spatial dimensions, while preserving a certain combination
of the two.
The configuration reminds us of the cholesteric phase of liquid crystals. 

Spatially modulated phases are known in condensed matter physics and
in QCD. In the Fulde-Ferrell-Larkin-Ovchinnikov phase, a Cooper
pair of two species of fermions with different Fermi momenta condenses
with non-vanishing total momentum \cite{FF,LO}. 
An analogous effect in QCD was studied in \cite{Alford}. 
It has also been shown that finite density QCD in the large $N_c$
limit is unstable against forming the chiral density wave \cite{Rubakov,Son}. 
Gravity theories with Chern-Simons terms may provide dual descriptions
of such systems and clarify aspects of their phase transitions. 

For example, the Brazovskii model \cite{Brazovskii} generates
a spatially modulated phase, and it has been applied 
to a variety of physical problems \cite{Hohenberg}. In this model, 
a non-standard dispersion relation is postulated 
so that the fluctuation spectrum has a minimum 
at non-zero momentum. The gravity theory discussed in this paper
provides a holographic realization of  
a similar dispersion relation.

In this paper, we use the Maxwell theory with the Chern-Simons term 
coupled to the gravity in $AdS_5$ as 
a phenomenological 
model of quantum critical phenomena in the spirit of 
\cite{Hartnoll,Herzog}. 
To have an explicit description of the
field content and interactions of the dual field theory, 
we need to identify a specific superstring construction where
the instability takes place. We examined the simplest case of 
the three-charge black hole in the 
type IIB superstring theory on $AdS_5 \times S^5$
and found that the Chern-Simons coupling of the low energy gravity
theory barely satisfies the stability bound. More specifically, 
when the three charges are the same, the effective Chern-Simons
coupling $\alpha$ is only 0.4 $\%$ less than 
the critical value $\alpha_{{\rm crit}}$ for the instability. 
There is a limit of an extreme ratio of charges, where
an effective $\alpha$  coincides
with $\alpha_{{\rm crit}}$ and the black hole
becomes marginally stable. 

This seems to indicate that, if we survey a wider class of examples, 
we may be able to find a theory with a Chern-Simons coupling large enough 
to cause an instability.
Generally speaking, the Chern-Simons coupling for a gauge field 
in $AdS_5$ is proportional to the chiral anomaly of the corresponding 
current in the dual conformal field theory \cite{Witten}. In particular, 
for the type IIB superstring theory on $AdS_5$ times a toric 
Sasaki-Einstein manifold, the Chern-Simons coupling is determined by the
toric data, or equivalently by the combinatorial data of the quiver
diagram for the dual gauge theory \cite{Tachikawa}. It would be interesting to
find an explicit example where the Chern-Simons coupling exceeds the
stability bound. Or, one may try to prove that such theories
are all in the Swampland \cite{SwampOne,SwampTwo}. 

Even for a theory with $\alpha < \alpha_{{\rm crit}}$, in which
the Chern-Simons term is not strong enough to cause instability, 
the non-standard dispersion relation is
noteworthy by itself with potential applications to physical problems. 
For example, a plasmino in QCD 
is a collective mode of quarks whose spectrum has a minimum 
at a non-zero momentum \cite{Klimov:1981ka,Weldon:1982bn}. 
In the presence of a plasmino, the dilepton production rate diverges 
at the minimum of the spectrum due to the Van Hove singularity,
$i.e.$, the divergence of the density of states per unit 
energy \cite{Braaten:1990wp,LeBellac}. 

We should also point out that another type of instability of rotating
charged black holes was suggested in \cite{gmt,kunz}.
While the Chern-Simons term seems to play a role there, 
we have found no obvious connection to the instability
discussed in this paper. Effects of the bulk Chern-Simons terms
on hydrodynamics of the dual field theories have been studied in 
\cite{Erdmenger,Dutta,Torabian,DSon}.
In \cite{Matsuo,Sahoo}, dispersion relations of hydrodynamic waves
in the Reissner-Nordstr\"om geometry with the Chern-Simons term
are discussed. Since the authors of 
these papers relied on power series expansions around $k=0$, 
they did not observe the instability we found in this paper
since the range of instability is away from $k=0$
as shown in Fig. 2. 

This paper is organized as follows. In section 2, we show that the
five-dimensional Maxwell theory with the Chern-Simons term is unstable
in the presence of a constant electric field. The metric is treated as 
non-dynamical in this analysis. In section 3, we turn on the metric fluctuation
and study the stability of the near-horizon geometry of the extremal 
Reissner-Nordstr\"om black hole in $AdS_5$. In section 4,
we generalize the analysis of section 3 to the full 
Reissner-Nordstr\"om solution. 
We solve the linearized equations around the black hole geometry
 and identify the critical temperature $T_{{\rm crit}}$
of the phase transition. 
We examine the onset of the phase transition and interpret the 
result from the point of view of the dual field theory. 
In section 5, we show that the three-charge black hole in the type IIB
superstring theory on $AdS_5\times S^5$ is barely stable against
the type of instability discussed in this paper. 

\section{Maxwell Theory with Chern-Simons Term}\label{S:MaxwellCS}

It is well-known that the three-dimensional Maxwell theory with the Chern-Simons
term is massive \cite{Deserone,Desertwo}. The equation of motion for the
2-form field strength $F$ is given by
\beq
d^* F + \alpha F = 0, \label{three}
\eeq
where $\alpha$ is the Chern-Simons coupling constant. 
Applying $d^*$ to this equation
and using the Bianchi identity $d F = 0$, one finds
$$ \Box F= d^*d^* F = - \alpha d^* F = \alpha^2 F.$$
Thus, the Chern-Simons term in three dimensions induces 
the mass $|\alpha|$ of the gauge field. 

Surprisingly, we find that the Chern-Simons term in five dimensions can turn 
the Maxwell theory tachyonic.  In this section,
we will demonstrate this by treating gravity as non-dynamical. 
Coupling to gravity will be
studied in the following sections.  Consider the following Lagrangian density,
\beq
{\cal L} =  -\frac{1}{4} \sqrt{-g} F_{IJ} F^{IJ} 
+ \frac{\alpha}{3!} \epsilon^{IJKLM} A_I F_{JK} F_{LM}, \label{fivelag}
\eeq
with the equation of motion,
\beq
\partial_J(\sqrt{-g} F^{JI}) + \frac{\alpha}{2}
\epsilon^{IJKLM} F_{JK} F_{LM} = 0.\label{five}
\eeq
We use the
almost positive convention for the metric $g_{IJ}$ $(I,J=0,...,4)$.
Choose a background solution $F^{(0)}$ and linearize \eqref{five} around it
by substituting $F = F^{(0)} + f$ in \eqref{five}. The linearized equation for
$f$ is given by 
\beq
\partial_J(\sqrt{-g} f^{JI}) + \alpha
\epsilon^{IJKLM} F^{(0)}_{JK} f_{LM} = 0.\label{linear}
\eeq
If $F^{(0)}$ is magnetic, this equation is similar to 
\eqref{three}; the fluctuation $f_{IJ}$ is massive and the 
configuration is stable. 
If $F^{(0)}$ 
is electric, on the other hand,
\eqref{linear} has tachyonic modes as we now explain.

Suppose the five-dimensional space is flat $\bR^{1,4}$,
regard it as the product $\bR^{1,1} \times \bR^3$, and use coordinates 
$(x^{\mu=0,1}, y^{i=2,3,4})$. Let us turn on a constant electric field in 
the $x^1$ direction, 
\begin{align}
 F^{(0)}_{\mu\nu}& = E\epsilon_{\mu\nu},\nonumber \\
 F_{\mu i}^{(0)}&=0, ~~~ F_{ij}^{(0)}=0. \label{volume}
\end{align}
The equation of motion \eqref{linear} is then,
\begin{align}
 & \partial^\mu f_{\mu \nu} + \partial^i f_{i\nu} = 0,
\nonumber \\
& \partial^\mu f_{\mu i} + \partial^j f_{ji} 
- 2\alpha E \epsilon_{ijk} f_{jk} = 0.  
\end{align}
Our $\epsilon$-symbol convention is such that 
$\epsilon_{01}=1$ and $\epsilon_{234}=1$.
By multiplying $\epsilon_{ijk} \partial_j$ to 
the second equation, we obtain
\beq
\left(\partial^\mu \partial_\mu + \partial^j \partial_j \right) f_i
 -4 \alpha E \epsilon_{ijk} \partial_j f_k = 0, \label{eqforf}
\eeq
where
$$ f_i = \frac{1}{2} \epsilon_{ijk} f_{jk}.$$
To derive \eqref{eqforf}, we used 
the Bianchi identities,
$$  \partial_i f_{\mu j} - \partial_j f_{\mu i} = 
\partial_\mu f_{ij}, ~~ \epsilon_{ijk} \partial_i f_{jk} =
2 \partial^i f_i = 0. $$

In the momentum basis $e^{ip_\mu x^\mu+ ik_iy^i}$, the operator 
$\epsilon_{ijk} \partial_j$ has eigenvalues $\pm k$ and $0$,
where $k =|\vec{k}|$.
However, the eigenvalue $0$ corresponds to $f_i \sim k_i$, 
which is prohibited by the Bianchi identity $k^i f_i = 0$. 
Thus, the linearized equation \eqref{eqforf} gives the dispersion 
relation,\footnote{At this point, we should note that there is a
similarity of this dispersion relation to eq. (17) of \cite{Domokos}
if we set $m_\rho = m_{a_1}$ in the paper and interpret
$m_\rho^2$ as being equal to $(p_1)^2$.}
\begin{align}
(p_0)^2 - (p_1)^2 &= 
k^2 \pm 4 \alpha E k \nonumber \\
& = (k \pm 2\alpha E)^2 - 
  4 \alpha^2 E^2. \label{mass}
\end{align}
We find tachyonic modes in $\bR^{1,1}$ in the range of
$0 < k < 4 |\alpha E|$. 

It is instructive to compare this with the case when we turn on 
a constant magnetic field,
\begin{align}
  F_{34}^{(0)}& =- F_{43}^{(0)}= B, \nonumber \\
  F_{IJ}^{(0)}&=0~~({\rm otherwise}). 
\end{align}
By repeating the previous analysis, we find the dispersion relation,
$$ (p_0)^2 - (p_1)^2 - (k_2)^2 
= \left( \sqrt{ (k_3)^2 + (k_4)^2 + 4 \alpha^2 B^2}
     + 2 |\alpha B| \right)^2. $$
In particular, when $k_3=k_4=0$, the equation gives
$p_0^2 - p_1^2 - k_2^2= ( 4 \alpha B)^2$, reproducing
the topologically massive gauge field in three dimensions. 

In the following sections, we 
will examine stability of the extremal Reissner-Nordstr\"om black hole
in $AdS_5$. If the boundary theory is on $\bR^{1,3}$, 
the near-horizon geometry of an extremal black hole
takes the form $AdS_2 \times \bR^3$ with an electric field proportional
to the volume form of $AdS_2$.  In such a configuration, 
the effective mass squared in $AdS_2$ is again given by the
right-hand side of  \eqref{mass}. 
The configuration is unstable if $-4 \alpha^2 E^2$ violates the
Breitenlohner-Freedman bound $m_{BF}^2$ in $AdS_2$,  namely,
\beq
    4 \alpha^2 E^2 > | m_{BF}^2| =  \frac{1}{4r_2^2}, \label{inequality}
\eeq
where $r_2$ is the curvature radius of $AdS_2$.
If this inequality is satisfied, 
the instability happens for non-zero momenta in the range,
\beq
2|\alpha E|\left( 1 -\sqrt{1
- \frac{1}{16\alpha^2E^2r_2^2}}\right) < k <
2|\alpha E|\left( 1 + 
\sqrt{1 - \frac{1}{16\alpha^2 E^2 r_2^2}}\right).\label{momentum}
\eeq
It is interesting to note that the zero momentum $k=0$ is excluded from
the instability range. Thus, the condensate of the gauge field happens for 
non-zero momentum in the $\bR^3$ direction of the near-horizon geometry. 

As we shall see in the next section, the value of $\alpha$ for 
the minimal gauged supergravity is such that
$4 \alpha^2 E^2$ exceeds the stability bound as in
\eqref{inequality}. 
This, however, does not mean that
extremal charged black holes in the minimal gauged supergravity are unstable 
since we must take into account the coupling of the Maxwell field 
to other degrees of freedom in the supergravity theory. We will perform this
analysis in the next section. 

\section{Coupling to Gravity}\label{CouplingToGravity}

The background electric field causes mixing
of the gauge field with the metric at the quadratic order, and
it modifies the stability condition. In this section, we
will study stability of the near-horizon geometry of the 
extremal Reissner-Nordstr\"om solution in $AdS_5$. It is
a solution to the Maxwell theory with the Chern-Simons term coupled 
to the Einstein gravity with negative cosmological constant, 
\beq
16\pi G_5 {\cal L} = \sqrt{-g}\left( R + \frac{12}{\ell^2} 
- \frac{1}{4} \ell^2 F_{IJ}F^{IJ}\right)
 + \frac{\alpha}{3!}  \ell^3 \epsilon^{IJKLM} A_I F_{JK} F_{LM}. \label{minimal}
\eeq
The curvature radius $r_5$ of the $AdS_5$ solution
in this theory is equal to $\ell$. In the following, we will work 
in the unit of $\ell = 1$. 
This is also the Lagrangian density of the minimal gauged supergravity 
in five dimensions \cite{Gunaydin}. In this case, supersymmetry determines
the Chern-Simons coupling $\alpha$ as
\beq
\alpha = \frac{1}{2\sqrt{3}}. \label{coupling}
\eeq
In this and next sections, we will treat \eqref{minimal} as a phenomenological 
Lagrangian with $\alpha$ as its parameter. 

\subsection{$AdS_2 \times \bR^3$}

Let us first consider the extremal black hole solution which is
asymptotic to $AdS_5$ in the Poincar\'e coordinates. It describes
the dual conformal field theory on $\bR^{1,3}$ with non-zero
chemical potential and at zero temperature. 
The near-horizon geometry of the extremal 
black hole is $AdS_2 \times \bR^3$ with the metric 
\beq
  ds^2 =  \frac{-(dx^0)^2 + (dx^1)^2}{12(x^1)^2}
              + d\vec{y}^2,~~~\vec{y}=(y^2, y^3, y^4).
 \label{nearhorizon}
\eeq
Note that the curvature radius $r_2$ of $AdS_2$ is $1/\sqrt{12}$;
the curvature is stronger near the horizon. 
The electric field strength near the horizon is proportional 
to the volume form of $AdS_2$ and is given by
\beq 
  F_{01}^{(0)} = \frac{E}{12 (x^1)^2}, ~~ E= \pm 2\sqrt{6}.
\eeq
For the minimal gauged supergravity with $\alpha$ given by \eqref{coupling},
\beq 
 4\alpha^2 E^2 = 8 > |m_{BF}^2| = \frac{1}{4r_2^2} =  3.
\eeq
Thus, if gravity is treated as non-dynamical, the gauge field fluctuation
near the horizon violates the Breitenlohner-Freedman bound for this value of $\alpha$.

We decompose the metric $g_{IJ}$ into the background $g_{IJ}^{(0)}$ and the fluctuation $h_{IJ}$ as $g_{IJ}=g_{IJ}^{(0)}+h_{IJ}$. The indices are raised/lowered by using the background metric. Notice that $g^{IJ}=g^{IJ(0)}-h^{IJ}+O(h^{2})$ so that $g^{IJ}g_{JK}=\delta^I_{~K}$.
In the presence of the background electric field $F^{(0)}_{\mu\nu}$, 
the unstable gauge field components $f_{\mu i}, f_{ij} \neq 0$ 
mix with the off-diagonal elements $h_\mu^{~i}$ of the metric perturbation 
through the gauge kinetic term,
\beq
 F_{IJ} F^{IJ} = 4 F^{(0)\mu\nu} h_\mu^{~i} f_{\nu i} + \cdots. \label{kinetic}
\eeq
Thus, in the stability analysis, we have to take into account the mixing.
One can think of $h_\mu^{~ i}$ as the Kaluza-Klein gauge field upon 
reduction on $\bR^3$. Since we are considering a sector with non-zero
momentum $\vec{k}$ along $\bR^3$, the Kaluza-Klein gauge field on $AdS_2$
has mass $\vec{k}^2$. 

To examine the stability of the black hole solution,
we can apply the standard linear perturbation theory.
In the present situation, however, there is a simpler way 
as we describe here. Suppose that the momentum $\vec{k}$ on $\bR^3$ is 
in the $y^2$ direction.
To derive the effective action
for the Kaluza-Klein gauge field $h_\mu^i$ in $AdS_2$, 
it is convenient to 
reduce the Einstein 
action in \eqref{minimal} along the $y^{3,4}$ directions first.
This gives rise to two gauge fields $(h_{\mu}^{~i}, h_2^{~i})$ ($i=3,4$) on
$AdS_2 \times \bR_{y^2}$, with the effective Lagrangian
\begin{align}
& \sqrt{-g_{5d}^{(0)}} \left( R + 12\right) \nonumber \\
&\rightarrow
\sqrt{-g_{3d}^{(0)}}
\left[ - \sum_{i=3,4}\left(\frac{1}{4} K_{\mu\nu}^{i} K^{i \mu\nu}
       + \frac{1}{2} K_{\mu 2}^{i}K^{i\mu 2} \right) + ({\rm terms~not~involving}
~ h_\mu^{~i}, h_2^{~i}) \right], 
\end{align}
where the gauge field strengths are
$$ K_{\mu \nu}^{i} = \partial_\mu h_\nu^{~i} - \partial_\nu h_\mu^{~i}, 
~ K_{\mu 2}^{i} = \partial_\mu h_2^{~i} - \partial_2 h_\mu^{~i} ~~
(\mu, \nu= 0,1; ~i=3,4). $$
Upon further reduction in the $y^2$ direction with momentum $k$, 
the effective Lagrangian density for the Kaluza-Klein gauge field
is
\beq
{\cal L}_{eff}
= - \sqrt{-g_{2d}^{(0)}}\sum_{i=3,4}
\left[\frac{1}{4} K_{\mu\nu}^{i} K^{i \mu\nu}
       + \frac{1}{2}  \left|\partial_\mu h_2^{~i}- i k h_\mu^{~i} \right|^2
     \right]. 
\label{Stuckelberg}
\eeq
We see that the off-diagonal elements $h_\mu^{~i}$ ($i=3,4$) of
the metric fluctuation give rise to two massive gauge fields of mass
$|k|$ on $AdS_2$ with
$h_2^{~i}$ serving as the requisite St\"uckelberg fields. 

Let us dualize the Kaluza-Klein field strength $K_{\mu\nu}^i$ on $AdS_2$ and
write it as a function
$K_i$ times the volume form,
$$ K_{01}^i =  \frac{K_i}{12 (x^1)^2}.$$
 The equations of motion for 
$f_i = \frac{1}{2} \epsilon_{ijk} f_{jk}$ and $K_j$ are
derived from the Lagrangian density which is
\eqref{fivelag}
plus \eqref{Stuckelberg} with the coupling \eqref{kinetic}.
They can be organized into the form,
\begin{align}
 & \left(\Box_{AdS_2} + \partial^j \partial_j \right) f_i
 -4 \alpha E \epsilon_{ijk} \partial_j   f_k 
 +E \epsilon_{ijk} \partial_j  K_k = 0, \nonumber \\
& E\ \Box_{AdS_2} f_i
+ \left(\Box_{AdS_2} + \partial^j \partial_j \right)
\epsilon_{ijk} \partial_j  K_k = 0 . \label{double}
\end{align}
The effective mass $m$ of these fields in $AdS_2$ can then be computed 
by solving
\beq
  {\rm det}\left(\begin{array}{cc}
m^2 - k^2 - 4\alpha E k&
E \\
Em^2 & m^2 - k^2
\end{array}
\right) = 0, \label{determinant}
\eeq
where $k=\pm |k|$. We find
\beq
  m^2 = \frac{1}{2}\left( 2k^2 + E^2 + 4\alpha  E k
 \pm \sqrt{E^4 +8 \alpha E^3 k 
+ 4(1 + 4 \alpha^2) E^2 k^2}\right). \label{massspectrum}
\eeq    

Minimizing $m^2$ with respect to $k$ and choosing the minus
sign in \eqref{massspectrum}, we obtain
the lowest value of $m^2$ as
\begin{align}
m^{2}_{\mbox{\scriptsize min}}=\frac{E^2 \left(-64 \alpha ^6-24 \alpha ^4+6
   \alpha ^2-\left(16 \alpha ^4+4 \alpha
   ^2+1\right)^{3/2}+1\right)}{2 \left(4 \alpha
   ^2+1\right)^2}.
\nonumber
\end{align}
Substituting $E = 2 \sqrt{6}$ for the near-horizon geometry, 
we find numerically that the lowest value of $m^2$ violates the
Breitenlohner-Freedman bound if 
\beq
  |\alpha| > \alpha_{{\rm crit}}= 0.2896 \cdots . \label{lowestalpha}
\eeq

The value of $\alpha$ for the minimal gauged supergravity
is 
$$ \alpha = \frac{1}{2\sqrt{3}}=0.2887 \cdots . $$
Thus, the supergravity theory is stable against the fluctuation of the
gauge field, but barely so  (with a margin less than 0.4 $\%$). 

\subsection{$AdS_2 \times S^3$}

For completeness, let us consider the case when the boundary theory is
on $\bR \times S^3$. The near-horizon geometry is $AdS_2 \times S^3$.
Let us denote the curvature radii of $AdS_2$ and $S^3$ by 
$r_2$ and $r_{3}$, respectively. They are related to the electric field
strength $E$ and the cosmological constant $\Lambda$, which is $-6$ in the $AdS_2\times \mathbb{R}^3$ limit, by 
\begin{align}
   & \Lambda  = -\frac{1}{2r_2^2} + \frac{2}{r_3^2}, \nonumber \\
   & E^2  = \frac{2}{r_2^2} + \frac{4}{r_3^2}.
\end{align}
Note that, in the limit $r_3 \rightarrow \infty$, where $S^3$ becomes
$\bR^3$, this reproduces $E= \pm 2 \sqrt{6}$ in our unit. 

As in the previous case, we consider fluctuations of the metric
$g_{IJ} = g_{IJ}^{(0)} + h_{IJ}$ and the gauge field
$F_{IJ}= F_{IJ}^{(0)} + f_{IJ}$ from their classical values indicated
by ${}^{(0)}$. We expand the Einstein equation,
\beq
  R_{IJ} - \frac{1}{2} g_{IJ} R = \frac{1}{2}\left(
 F_{IK}F_J^{~K} - \frac{1}{4} g_{IJ} F_{KL}F^{KL}\right),
\eeq
and the Maxwell equation modified by the Chern-Simons term,
\beq
 \sqrt{-g} \nabla_J F^{JI} +\frac{\alpha}{2} \epsilon^{IJKLM} F_{JK}F_{LM} = 0,
\eeq
to the linear order in $h_{IJ}$ and $f_{IJ}$. 

The linearized equations
for $f_{ij}$ and $K_i$, where $K_i$ is defined such that $K_i ({\rm vol~AdS}_{2})_{\mu\nu} = 2 \nabla_{[ \mu} h_{\nu]i}$,
can be written as,
\begin{align}
&  \left(\Box_{AdS_2} + \Delta_{S^3}\right) f
- 4\alpha E d^* f + E dK = 0 , \nonumber \\
& E \Box_{AdS_2} f 
+ \left(\Box_{AdS_2} + \Delta_{S^3} +\frac{4}{r_3^2} \right)
dK = 0.
\end{align}
These equations are similar to \eqref{double}, except for the last term
$\frac{4}{r_3^2}$ in the second equation.  
Here ${}^*$ means the Hodge dual on $S^3$. Since $d^*$ is hermitian 
when acting on the space of two-forms on $S^3$, decompose $f$ into
its eigenstate. Its eigenvalue is known to be $k=\pm (n+2)/r_3$, where
$n=0,1,2,...$. Since $\Delta_{S^3}= -(d^*)^2$ when acting on
$f$ satisfying the Bianchi identify $df=0$, we can set $\Delta_{S^3}=-k^2$. 

The mass $m$ on $AdS_2$ then satisfies the determinant equation,
\beq
  {\rm det}\left(\begin{array}{cc}
m^2 - k^2 - 4\alpha E k&
E \\
Em^2 & m^2 - k^2 + 4 /r_3^2
\end{array}
\right) = 0. 
\eeq
This can be solved to obtain,
\beq
m^2 = \frac{1}{2}\left[
2k^2 + E^2 +4\alpha E k - \frac{4}{r_3^2} \pm
\sqrt{E^4 + 8\alpha E^3 k + \frac{16}{r_3^4}
+ \frac{32 \alpha k}{r_3^2}
+ 4 E^2\left(k^2 + 4 \alpha^2 k^2 - \frac{2}{r_3^2}\right)}\right]. \label{massspectra}
\eeq
In the limit of $r_3 \rightarrow \infty$, this reduces
to the previous result \eqref{massspectrum}. 

We have numerically checked that, for a wide range of $\Lambda$ and $E$, 
the Breitenlohner-Freedman bound is not violated in the minimal gauged
supergravity, where $\alpha = \frac{1}{2\sqrt{3}}$. It is interesting to 
note that, in the limit of $\Lambda \rightarrow 0$ but with non-zero $E$, 
the lowest $m^2$ in \eqref{massspectra} saturates the Breitenlohner-Freedman 
bound \cite{fujii}, which is
\beq -\frac{1}{4r_2^2} = -\frac{1}{12}(E^2 - 2 \Lambda) = 
  -\frac{E^2}{12}.
\eeq

\section{Phase Transition and Critical Temperature}\label{S:PhaseTransition}

In the last section, we studied the instability of the 
near-horizon region of the extremal Reissner-Nordstr\"om solution.
This gives a sufficient condition for the solution to be unstable.
However, as we will see in this section, the condition turns out
to be not necessary. To clarify the nature of the phase transition
and identify the critical temperature, we study linear perturbation
to the full Reissner-Nordstr\"om black hole in $AdS_5$.

\subsection{Geometry and Equations}

The Reissner-Nordstr\"om black hole has the metric
\begin{equation}\label{E:RNblackhole}
ds^2=-H(r) dt^2 + \frac 1 {H(r)} dr^2 + r^2 d{\vec y}^2\;, \qquad \vec y = 
(y^2, y^3, y^4) \;.
\end{equation}
Note $\sqrt{-g^{(0)}}=r^3$.
The gauge field strength is given by
\begin{equation}
\qquad F^{(0)}=\frac{Q}{r^3} dt\wedge dr\;.
\end{equation}
The function $H(r)$ is given by
\begin{equation}
H(r)=r^2\left[1-\left(1+\frac{\mu^2}{3r_+^2}\right)\left(\frac {r_+} r\right)^4+\frac{\mu^2}{3r_+^2}\left(\frac{r_+} r\right)^6\right]\;,
\end{equation}
where $Q=-2 \mu r_+^2$.

The equation of motion coming from the variation of the gauge field $a_i$ is
\begin{equation}
\partial_{\mu} (\sqrt{-g^{(0)}} f^{\mu i})+\partial_j (\sqrt{-g^{(0)}} f^{ji}) - 2\alpha \frac{Q}{r^3} \epsilon_{ijk} f_{jk}-\partial_{\rho} 
\left(\sqrt{-g^{(0)}} \frac {Q}{r^3} \epsilon^{\mu\rho} h_{\mu}^{~i}\right)=0\;.
\end{equation}
In the black hole background \eqref{E:RNblackhole}, it becomes
\begin{equation}\label{E:aieq}
-\frac{r}{H(r)} \partial_t f_{ti}+\partial_r ( r H(r) f_{r i})+ \frac 1 {r} \partial_j f_{ji}-2\alpha \frac {Q}{r^3} \epsilon_{ijk} f_{jk}+Q K^i=0\;,
\end{equation}
where $K^i=\partial_t h_r^{~i} - \partial_r h_t^{~i}$.
By operating $\epsilon_{ijk} \partial_j$ on this equation, we obtain
\begin{equation}\label{E:EOM1}
-\frac{r}{H(r)} \partial^2_t f_i + \partial_r ( H(r) r \partial_r f_i) + \frac 1 r \Delta_{\bR^3} f_i - 4\alpha \frac{Q}{r^3} \epsilon_{ijk} \partial_j f_k + Q \epsilon_{ijk} \partial_j K^k=0\;,
\end{equation}
where $f_i = \frac 1 2 \epsilon_{ijk} f_{jk}$ and 
$\Delta_{\bR^3} = \partial_{y^2}^2 + \partial_{y^3}^2 + \partial_{y^4}^2$.

To obtain the equation of motion that comes from the variation of the off-diagonal metric elements, let us use the Kaluza-Klein reduction in 
the presence of momentum $\vec{k}$ along the $y^2$ direction.
The effective Lagrangian has the form
\begin{equation}
\mathcal{L}_{eff}=-r^3 \sqrt{-g_{2d}^{(0)}} \sum_{i=3,4} \left[ \frac 1 4 r^2 K^i_{\mu\nu} K^{i\mu\nu} + \frac 1 2 \left|\partial_{\mu} h_2^{~i} - ik h_{\mu}^{~i}\right|^2
\right]\;.
\end{equation}
The $r^3$ factor comes from the volume form on the
$\bR^3$  directions with coordinates $\vec{y}$.
The equation of motion coming from the variation
with respect to the metric is given by
\begin{equation}
\partial_{\nu} (r^5 K^{\nu\mu i}) + r^3(-ik) (\partial^{\mu} h_2^{~i} - ik h^{\mu i}) - Q \epsilon^{\mu\rho} f_{\rho i}=0\;.
\end{equation}
Acting on the operator $\epsilon_{\alpha \mu}\partial_{\beta} g^{\alpha\beta(0)} \frac 1 {r^3}$, we can eliminate the term containing $\partial^{\mu} h_2^{~i}$.
Using $\epsilon_{\alpha\mu}\partial_{\nu}+\epsilon_{\mu\nu}\partial_{\alpha}+\epsilon_{\nu\alpha}\partial_{\mu}=0$ in two dimensions, we obtain
\begin{equation}
-\frac 1 2 \epsilon_{\mu\nu} \partial_{\beta} g^{\alpha\beta(0)} \frac 1 {r^3} \partial_{\alpha} r^5 K^{\nu\mu i}+k^2 K^i-Q\partial_{\mu}(g^{\mu\nu(0)} \frac 1 {r^3} f_{\nu i})=0\;.
\end{equation}
Further operating $\epsilon_{ijk}\partial_j$ on the equation,
\begin{equation}
\partial_{\beta}[ g^{\alpha\beta(0)} \frac 1 {r^3} \partial_{\alpha}( r^5 \epsilon_{ijk} \partial_j K^{k})]+\Delta_{\bR^3}^2 \epsilon_{ijk} \partial_j K^k + Q\partial_{\mu}(g^{\mu\nu(0)} \frac 1 {r^3} \partial_{\nu} f_{i})=0\;.
\end{equation}
More explicitly,
\begin{equation}\label{E:EOM2}
\left(-\frac 1 {r^3 H(r)} \partial_t^2 + \partial_r H(r)\frac 1 {r^3} \partial_r \right)\left( r^5 \epsilon_{ijk} \partial_j K^k + Q f_i\right) + \Delta_{\bR^3}^2 \epsilon_{ijk}\partial_j K^k=0\;.
\end{equation}

We have two sets of equations of motion \eqref{E:EOM1} and \eqref{E:EOM2}.
To simplify them, let us perform the following rescaling,
\begin{equation}
r\rightarrow \frac{r_+} u\;, \qquad t\rightarrow \frac{t}{r_+}\;,\qquad \vec{y}\rightarrow \frac{\vec{x}}{r_+} \;,
\end{equation}
and make the change of variables,
\begin{align}
&f_i(r)\rightarrow \phi(r), \nonumber \\
&\epsilon_{ijk} \partial_j K^k \rightarrow \frac{1}{\sqrt{3}r_+^2} u^3 \psi(r),
\end{align}
and set $q=\frac{\mu}{\sqrt 3 r_+}$.
The temperature $T$ is
\begin{equation}
T=\frac{r_+}{2\pi} \left(2-\frac{\mu^2}{3r_+^2}\right)\;.
\end{equation}
With the rescaled variables, the Reissner-Nordstr\"om black hole is
\begin{equation}
ds^2 = \frac{1}{u^2}\left( - \tilde{H}(u) dt^2 + \frac {1}{\tilde{H}(u)} du^2\right) + \frac{1}{u^2} d{\vec x}^2\;,
\end{equation}
where
\begin{equation}
\tilde{H}(u)=1-(1+q^2) u^4 + q^2 u^6\;.
\end{equation}
In these coordinates, the $AdS_5$ boundary is at $u=0$ and the black hole horizon is located at $u=1$.

Suppose that the fields $\phi$ and $\psi$ have time dependence $e^{-i \omega t}$.
Then the equations of motion for the fields $\phi$ and $\psi$ give
the following set of ordinary differential equations, 
\begin{equation}
\begin{split}
&\frac{\omega^2}{\tilde{H}(u)} \phi + u \partial_u \left(\tilde{H}(u)u^{-1} \partial_u \phi\right)- k^2 \phi + 8 \sqrt 3 \alpha q k u^2 \phi - 2 q u^2 \psi=0\\
&\frac{\omega^2}{\tilde{H}(u)} \left(\psi - 6 q u^2 \phi\right) + u^{-1} \partial_u \left[ \tilde{H}(u) u^3 \partial_u \left(u^{-2} (\psi - 6 q u^2 \phi)\right)\right]-k^2 \psi=0\;.
\end{split}
\end{equation}
Introducing a new function $\xi=\psi-6 q u^2 \phi$, the equations can be written as
\begin{equation}
\begin{split}
&\frac{\omega^2}{\tilde{H}(u)} \phi + u \partial_u(\tilde{H}(u)u^{-1}\partial_u \phi)-(k^2+8 \sqrt3 \alpha q k u^2 + 12 q^2 u^4)\phi - 2 q u^2 \xi=0\\
&\frac{\omega^2}{\tilde{H}(u)} \xi + u \partial_u(\tilde{H}(u)u^{-1}\partial_u \xi)-6 q k^2 u^2 \phi -(k^2-8u^2 -9u^2 q^2+12 q^2 u^4 )\xi=0\;.
\end{split}
\end{equation}
Interestingly, the two equations can be diagonalized by a $u$-independent matrix.
That is, for some linear combinations $\phi_1$ and $\phi_2$ of $\phi$ and $\xi$, we have
\begin{equation}\label{E:EOMnumeric}
\frac{\omega^2}{\tilde{H}(u)} \phi_i(u) + u \partial_u(\tilde{H}(u)u^{-1}\partial_u \phi_i(u))- \kappa_i(u) \phi_i(u)=0\;,
\end{equation}
where $i=1,2$ and 
\begin{equation}
\begin{split}
\kappa_i(u)=&k^2-4\sqrt 3 \alpha k q u^2 \\
&- 2 u^2 \left(2+q^2(2-6u^2)\mp \sqrt{4+4q^4-8\sqrt 3 \alpha k q -8\sqrt 3k q^3 \alpha+q^2(8+k^2(3+12\alpha^2))}\right)\;,
\end{split}
\end{equation}
where $\kappa_1$($\kappa_2$) chooses the minus (plus) sign on the 
right-hand side.
Our numerical analysis shows that only $\phi_2(u)$ can be an unstable mode.
It is related to the fact that, in the extremal limit, $\kappa_2$ corresponds 
to the smaller mass-squared in \eqref{massspectrum} in the near-horizon limit.

\subsection{Numerical analysis}

To solve the equations of motion \eqref{E:EOMnumeric} numerically, we impose the in-going boundary condition near the horizon $u=1$, and then evolve the solution to $u=0$, the $AdS_5$ boundary.
The asymptotic behavior of $\phi_i$ near $u=0$ is either $\phi_i \sim u^2$ or constant.
The former is normalizable and the latter is non-normalizable.
To find normalizable modes in the full Reissner-Nordstr\"om solution, we scan
the initial conditions and see when the fields vanish at $u=0$.

\begin{figure}[ht]
\centering
\includegraphics[width=7cm]{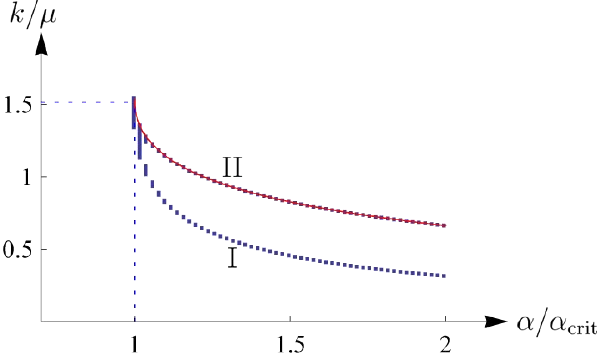}
\caption{For a given value of the Chern-Simons coupling $\alpha$, there is a discrete set of momenta $k$ for which
static solutions exist. The curves I and II indicate two of such momenta for each $\alpha$.
The red curve is the lower-end of the momentum range that violates the Breitenlohner-Freedman bound near the horizon.
Note that the red curve coincides with the curve II. However, there is 
another curve I with a lower momentum. This means
that the near-horizon analysis gives a sufficient but not necessary condition for the instability.
Both curves end at the same critical value of $\alpha$. 
 }\label{staticsolzeroT}
\end{figure}

First, let us consider the zero temperature limit ($q=\sqrt 2$)
and search for static solutions ($\omega=0$), which 
signal the onset of an instability.
The behavior of the fields near $u=1$ can be found from \eqref{E:EOMnumeric}
as 
\begin{equation}\label{E:zeroTstaticingoing}
\phi_i=(1 - u)^{-\frac 1 2 + \sqrt{\frac{\kappa_i(1)+3}{12}}}(1+\cdots)\;,
\end{equation}
where terms in $\cdots$ vanish at $u=1$.
In the actual numerical calculation in this section,
we include several subleading terms to improve accuracy.
For a given Chern-Simons coupling $\alpha$, static modes 
appear at discrete values of momentum $k$. The lowest two modes are plotted 
 in Fig. \ref{staticsolzeroT}.
As mentioned before, only the second field $\phi_2$ 
has normalizable static solutions.

The two curves in Fig. \ref{staticsolzeroT} are denoted as I and II. Both curves
terminate at $\alpha/\alpha_{{\rm crit}}=1$
and $k/\mu = 1.52 \cdots$.
The critical value of the Chern-Simons coupling 
$\alpha_{{\rm crit}} = 0.2896\cdots$ is the one we found
from the stability analysis of the near-horizon geometry
in the previous section. 
The curves are supposed to extend over $k/\mu = 1.52 \cdots$ and
come back to the right in a bell-shaped curves.
The upper branches of the curves
represent the upper bounds of unstable modes. 
However, we have not been able to plot them due to 
inaccuracy of our numerical computation. 

We also found out a static solution at zero momentum.
However, for this solution, the curl of the off-diagonal 
metric component $\epsilon_{ijk} \partial_j K^k$ is 
constant on $\mathbb{R}^3$.
This means that $K^i$ is linear in $\mathbb{R}^3$, and
the solution is not normalizable.

We note that the curve II fits with 
the red curve which is at the lower-end of 
the momentum range that violates the 
Breitenlohner-Freedman bound in the near-horizon 
$AdS_2 \times \mathbb{R}^3$ geometry.
As we saw in the previous section, the near-horizon
geometry is unstable in this momentum range, thus
the full Reissner-Nordstr\"om solution should also be
unstable. In fact the momentum range that violates 
the Breitenlohner-Freedman
bound is specified by $\kappa_2(1)<-3$, where
$\phi_2(u)$ oscillates infinitely many times as they approach 
the horizon as can be seen from \eqref{E:zeroTstaticingoing}.
On general ground, we expect an instability to occur in this range
\cite{Gibbons:2002pq}.

Interestingly, the instability condition $\kappa_2(1)<-3$ of the
near-horizon geometry is not necessary for the instability of 
the full solution.  
This is because there is yet another curve I, located outside of this 
momentum range. What happens is that the curve I corresponds to a normalizable
perturbation to the full Reissner-Nordstr\"om geometry, but
the corresponding mode becomes
non-normalizable in the near-horizon limit. 
Our numerical analysis shows that the critical Chern-Simons coupling 
$\alpha_{{\rm crit}}$ for
the curve I is the same as that for the curve II, even though the 
value of $\alpha_{{\rm crit}}$  was derived
from the near-horizon analysis. 

\begin{figure}[ht]
\begin{center}$
\begin{array}{cc}
\includegraphics[width=7cm]{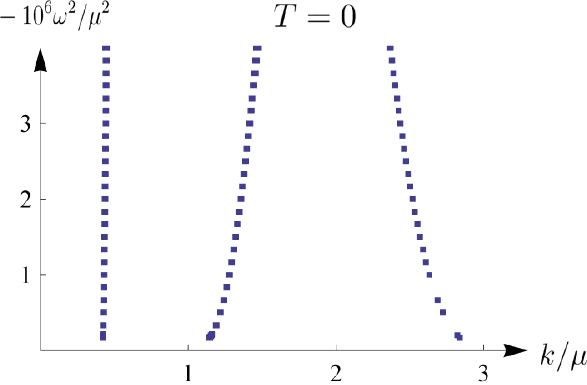} &
\includegraphics[width=7cm]{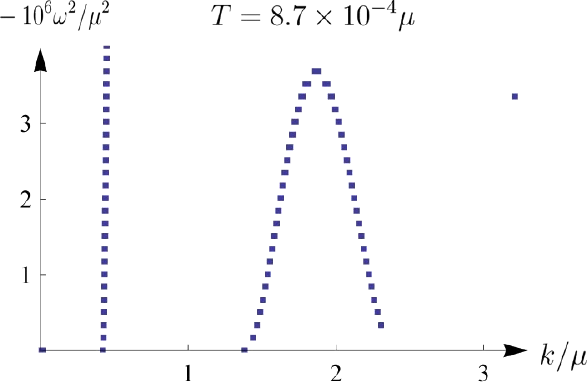}
\end{array}$
\end{center}
\caption{Left: Negative frequency squared as a function of momentum $k$ at zero temperature when $\alpha=1.6 \alpha_{\rm crit}$. Only positive $-\omega^2$ is plotted. The curves starting around 1 and 3 join to represent a tachyonic dispersion relation for the unstable mode predicted by the near-horizon analysis. The curve starting below 1 is also expected to be connected with another line in the higher momentum region to form a larger bell-shaped curve, but the large momentum part is difficult to analyze numerically. The zero momentum static solution does not extend to an unstable mode. Right: Negative frequency squared as a function of momentum at temperature $T=8.7\times 10^{-4} \mu$.}
\label{dispersionT}
\end{figure}

To see that these static solutions indeed signal instability, 
let us turn on $\omega$ with positive imaginary part in \eqref{E:EOMnumeric}.
We impose the in-going boundary condition, which is 
\begin{equation}
\phi_i = e^{-\frac{|\omega|}{12(1-u)}} (1-u)^{\frac 7 {36} |\omega|}\left(1+\cdots\right)
\end{equation}
in the zero temperature 
limit and 
\begin{equation}
\phi_i = (1-u)^{\frac {|\omega|}{4-2q^2}}\left(1+\cdots\right)\;.
\end{equation}
at a positive temperature.
Fig. \ref{dispersionT} shows the negative frequency squared as a function of the momentum $k$ at zero and finite temperature.
It shows that the upper and lower curves in Fig. \ref{staticsolzeroT} are boundaries of unstable modes.

The occurrence of instability by the Chern-Simons coupling is summarized concisely in Fig. \ref{criticaltemperature} and Fig. \ref{aTkgraph} in the introduction section of this paper.
For each Chern-Simon coupling $\alpha$, Fig. \ref{aTkgraph} shows an unstable region in the momentum-temperature plane. 
This is related to the curve I in Fig. \ref{staticsolzeroT}.
The range of unstable momenta never includes $k=0$.
The highest temperature with unstable modes is denoted as 
$T_{C} (\alpha)$. Fig. \ref{criticaltemperature} shows this critical temperature as a function of $\alpha$.
Below the critical temperature $T_{C} (\alpha)$, we expect an instability and the charge current gets a position dependent expectation value of the form of \eqref{VEV}.

\subsection{Spontaneous Current Generation}

The vacuum expectation value of the current $\vec{J}$ in the dual field theory
can be evaluated by extracting the asymptotic behavior of the corresponding
gauge field toward the boundary of $AdS_5$. In the absence of the 
Chern-Simons term, it is well-known that $\langle \vec{J} \rangle$ is given by
the normalizable part of $\sqrt{-g^{(0)}} f^{ri}$ evaluated at $r  \rightarrow
\infty$. The normalizable mode of $f_{ri}$ decays, but that effect is 
compensated by the scaling behaving of the metric so that we find a finite
limiting value in the low temperature phase. The Chern-Simons term gives 
rise to an additional term of the form $\alpha \mu 
\epsilon_{ijk} f_{jk}$. However, it vanishes at the boundary and does not
contribute to the expectation value. Thus, the vacuum expectation value
of the current in the low temperature phase is given by $\sqrt{-g^{(0)}} f^{ri}$ 
evaluated at the boundary of $AdS_5$. It takes the form,
\beq
   \langle \vec{J} (x) \rangle = {\rm Re}\left( \vec{u} e^{ikx}\right),
\label{expectation}
\eeq
where the polarization vector $\vec{u}$ obeys
\beq 
  \vec{k} \times  \vec{u} = \pm i |k|\vec{u}.
\eeq
From the analysis in the previous section, it is clear that
we should choose the plus (minus) sign when $\alpha E$ is positive
(negative). Namely, the sign of the Chern-Simons coupling determines whether
the circular polarization of the current expectation is clockwise or 
counter-clockwise. This configuration breaks 
translational and rotational symmetries, but a certain combination of the
two is preserved. 
The polarization of the current is helical and reminds us of the cholesteric
phase of liquid crystals. 

Since the gauge field mixes with the metric fluctuation $h_{\mu i }$ 
in the bulk, the corresponding component $T_{0i}$ of the energy-momentum
tensor has a non-zero expectation value at the boundary. This is expected
since the non-zero current in the spatial direction means that there is
a momentum density. 

\subsection{Spontaneous Breaking of Internal Symmetry}

So far, we have considered the case when the gauge group in the bulk
is $U(1)$.  Since
the $U(1)$ current commutes with itself, its expectation value does not
break the $U(1)$ global symmetry on the boundary. 

To realize spontaneous breaking of an internal symmetry, one possibility would
be to choose the gauge group to be non-abelian. 
The Chern-Simons term can be written in five dimensions if there
is a symmetric tensor $d_{abc}$ in the Lie algebra, such as in 
$SU(n)$ with $n \geq 3$. Suppose we turn on 
an electric field strength in a direction $T^a$ in the Lie algebra. 
According to \cite{GubserOne,GubserTwo},
the gauge kinetic term can generate instability in directions in the 
Lie algebra that do not commute with $T^a$. This breaks the symmetry
 homogeneously. On the other hand, the Chern-Simons 
term can cause a spatially modulated instability in directions
where $d_{abc}\neq 0$ with $T^a$. The competition of the two effects 
would be decided by the relative strength of the gauge coupling and
the Chern-Simons coupling. It would be interesting to study such 
an effect in a more explicit manner 
to identify the gravity dual of a spatially modulated phase with
spontaneous breaking of an internal symmetry.

\section{Three-Charge Black Holes in Type IIB Theory}

The consistent truncation of the type IIB theory on $AdS_5 \times S^5$
to the $U(1)^3$ gauged supergravity in five dimensions was given 
in \cite{Cvetic}. The bosonic action contains three gauge fields
for $U(1)^3$ and three scalar fields $X_1, X_2, X_3$ subject to 
the constraint $X_1 X_2 X_3 = 1$, in addition to the metric. 
This low energy theory admits the three-charge black hole solutions 
of \cite{Behrndt}. 
Here we will examine the stability of the near-horizon region of the three-charge black holes in the extremal limit.

\subsection{Case with Equal Charges}

Let us consider the case when the three charges are identical, which implies the scalar fields are constant $X_1 = X_2 = X_3 = 1$.
In this case, both the Lagrangian and the black hole configuration 
are symmetric under exchange of the three gauge fields $F_1, F_2, F_3$. 
It is convenient to take their linear combinations as
\begin{align}
  F &= \frac{1}{\sqrt{3}}\left( F_1 + F_2 + F_3 \right) \nonumber \\
  F_+ &= \frac{1}{\sqrt{3}}\left( F_1 + \omega F_2 + \omega^2 F_3 \right) 
\nonumber \\
F_- &= \frac{1}{\sqrt{3}}\left( F_1 + \omega^2 F_2 + \omega F_3 \right) .
\end{align}
They are eigenstates of the $\bZ_3$ permutation with eigenvalues $1$ and 
$\omega^{\pm 1}$, where $\omega= e^{2\pi i /3}$. 
In the black hole geometry, the $\bZ_3$
invariant gauge field $F$ has an electric component with $E = 2 \sqrt{6}$,
and $F_\pm = 0$. Similarly, 
fluctuations of the scalar fields from $X_1 = X_2 = X_3 = 1$
can be organized into eigenstates with eigenvalues $\omega^{\pm 1}$
under the $\bZ_3$ permutation.

The $\bZ_3$ invariant sector is the minimal gauged supergravity with
$\alpha = 1 /2\sqrt{3}$. To the quadratic order, the metric and the
$\bZ_3$ invariant gauge field do not mix with other fields. Thus, the
stability analysis with respect to them is exactly the same as
the one we performed in the previous section. The three-charge black hole
is barely stable in this sector, being within 0.4 $\%$ 
of the stability bound. 

Since the gauge fields $F_\pm$ have zero
expectation value on the black hole geometry, the $\bR^3$
components of these gauge fields do not couple with other degrees of
freedom in the quadratic order. It is convenient to write them as
$$ F_{\pm} = \frac{1}{\sqrt{2}}(f^{(1)} \pm i f^{(2)}). $$
With the standard normalization of their kinetic terms, the Chern-Simons term
takes the form, 
\beq
\frac{1}{8\sqrt{3}} \epsilon^{IJKLM}
F_{IJ} \left( a_K^{(1)} f_{LM}^{(1)} + a_K^{(2)} f_{LM}^{(2)}\right),
\eeq
where $a_I^{(i)}$ are the vector potentials for $f_{IJ}^{(i)}$ ($i=1.2$).
To the quadratic order, we can take the $\bZ_3$ invariant
$F_{IJ}$ to be its background value $F^{(0)}_{IJ}$. 

Since these gauge fields do not couple to other fields in the quadratic 
order, their linearized equations of motion are,
\beq
 \partial_J (\sqrt{-g} f^{(i)JI}) + \frac{1}{4\sqrt{3}} \epsilon^{IJKML}
F_{JK}^{(0)}f_{LM}^{(i)}=0,   ~~~(a=1,2).
\eeq
Comparing this with \eqref{linear}, we find $\alpha = \pm 1/4\sqrt{3}$. Since
$E = \pm 2 \sqrt{6}$ as in the previous example, the mass squared is given by
\beq
  -4 \alpha^2 E^2 = - 2.
\eeq
It is greater than the Breitenlohner-Freedman bound 
($m_{BF}^2=-3$ in our unit), 
and quadratic fluctuations in these gauge fields are stable.

\subsection{Case with Non-Equal Charges}

Next, let us consider the case when the three charges are different.
The five-dimensional Lagrangian is derived in \cite{Cvetic},
\begin{align}
&16\pi G_5 \mathcal{L} \nonumber\\
&=\sqrt{-g}\left(R- \frac 1 2 (\partial \phi_1)^2 - \frac 1 2 (\partial \phi_2)^2 + 4 \sum_a X_a^{-1} - \frac 1 4 \sum_a X_a^{-2} (F^a)^2\right) + \frac 1 4 \epsilon^{IJKLM} F^1_{IJ} F^2_{KL} A^3_{M}\;.
\end{align}
$X_a$ are functions of the two scalars $\phi_1$ and $\phi_2$ subject to the constraint $X_1 X_2 X_3=1$.
The Lagrangian admits $AdS_5$ black holes parametrized 
by three charges $q_1$, $q_2$ and $q_3$.
The metric is given by
\begin{align}\label{E:threechargemet}
&ds^2= -(H_1 H_2 H_3)^{-\frac 2 3} h(r) dt^2 + (H_1 H_2 H_3)^{\frac 1 3}\left(\frac {dr^2}{h(r)} + r^2 d\Omega_3^2\right) \nonumber\\
&H_a(r)= 1+ \frac{q_a}{r^2}\;,\qquad q_a=\mu\sinh^2\beta_a \;,\qquad a=1,2,3\nonumber \\
&X_a = H_a^{-1} (H_1 H_2 H_3)^{\frac 1 3} \nonumber \\
&h(r)=1-\frac{\mu}{r^2} + r^2 H_1 H_2 H_3 \nonumber \\
&A_a = (1-H_a^{-1})\coth \beta_a dt\;.
\end{align}
This is the metric whose foliating transverse space is $S^3$.
If it is $\mathbb{R}^3$ instead, the $S^3$ metric $d\Omega_3^2$ is replaced by the flat metric and $h(r)$ and $A^a$ are replaced with
\begin{align}
h(r)&=-\frac{\mu}{r^2} + r^2 H_1 H_2 H_3 \nonumber \\
A_a &= \frac{1-H_a^{-1}}{\sinh \beta_a} dt\;.
\end{align}
Given the charges $q_a$, it may be possible to choose $\mu$ such that the black hole becomes extremal. That is, the inner and the outer horizons coincide.
For the extremal case, the near-horizon geometry is $AdS_2\times S^3$ or $AdS_2\times\mathbb{R}^3$: if the horizon occurs at $r=r_0$, for small $\rho=r-r_0$, $h(r)=\frac 1 2 h''(r_0) \rho^2$.
Hence the geometry becomes
\beq
ds^2=\frac 1 {a_1}\left[- \rho^2 dt^2 + \frac{d{\rho}^2}{\rho^2}\right]+\frac{1}{a_2} d\Omega_3^2\;,
\eeq
where $a_1=\frac 1 2 (H_1 H_2 H_3)^{-\frac 1 3} h''$, and $a_2=(H_1 H_2 H_3)^{-\frac 1 3}r_0^{-2}$ for $S^3$ 
and $a_{2}^{-1} d\Omega_3^2$ is replaced with the flat metric for $\mathbb{R}^3$.
$H_a$ and $h''$ are implicitly evaluated at $r=r_0$.

We want to analyze the linear fluctuations near the horizon in the extremal limit.
Let $F_a = F^{(0)}_a+ f_a = F^{(0)}_a+ da_a$ and $g_{IJ}=g^{(0)}+h_{IJ}$.
If we focus on the fluctuations of the $a_i$ and $h_{\mu i}$ fields only, we find that the linear fluctuations of the scalar fields $\phi_1$ and $\phi_2$ do not couple to them.
Therefore, we may use the background value of the scalar fields.
In this case, we can derive the equations of motion as in the previous case, and the result is
\begin{align}
&(\Delta_2 + \Delta_3) f^1 -X_1^2  E_3 d^*f^2 - X_1^2 E_2 d^* f^3 +E_1 d K=0 \nonumber\\
&- X_2^2 E_3 d^*f^1 +(\Delta_2 + \Delta_3) f^2  - X_2^2 E_1 d^* f^3+E_2 d K=0 \nonumber\\
& - X_3^2 E_2 d^*f^1 - X_3^2 E_1 d^* f^2+(\Delta_2 + \Delta_3) f^3 +E_3 d K=0 \nonumber\\
&\frac{E_1}{X_1^2} \Delta_2 {}^*f^1+\frac{E_2}{X_2^2} \Delta_2 {}^*f^2+\frac{E_3}{X_3^2} \Delta_2 {}^*f^3+(\Delta_2+\Delta_3+4a_2) dK=0\;.
\end{align}
$E_a$ are the electric fields such that $dA_a = E_a (H_1 H_2 H_3)^{-\frac 1 6} dt \wedge dr$.
The above four equations give a mass matrix equation
\beq
\det\begin{pmatrix} m^2 - k^2 & -E_3 X_1^2 k & -E_2 X_1^2 k & E_1 \\
											-E_3 X_2^2 k & m^2 -k^2 & -E_1 X_2^2 k & E_2 \\
											-E_2 X_3^2 k & -E_1 X_3^2 k & m^2 -k ^2 & E_3\\
											\frac{E_1}{X_1^2} m^2 & \frac{E_2}{X_2^2} m^2 & \frac{E_3}{X_3^2} m^2 & m^2-k^2 + 4a_2\end{pmatrix}=0\;. \label{massmatrix}
\eeq
Solving this equation for $m^2$, we obtain the mass spectrum.

When two of the three charges are the same, we can analyze the mass spectrum analytically for the $AdS_2\times \mathbb{R}^3$ geometry.
In this case, only the ratio of the charges matter.
Let the charge assignments be $(q_a)=(1,q,q)$.
Demanding $f(r_0)=f'(r_0)=0$ at some $r=r_0$, we obtain the relation $q=x(2x+1)$ and $\mu=4x(1+x)^3$ where $x=r_0^2$.
Let us parametrize the extremal solutions in terms of $x$.
Then the various functions at the horizon are given by
\begin{align}
&H_1=x^{-1}(1+x)\;,\qquad H_2=H_3=2(1+x)\;, \qquad H_1 H_2 H_3 = 4 x^{-1} (1+x)^3\\
&X_1=2^{\frac 2 3} x^{\frac 2 3}\;,\qquad X_2=X_3=2^{-\frac 1 3} x^{-\frac 1 3}\\
&E_1=2^{\frac 7 3} x^{\frac 5 6}\;,\qquad E_2=E_3=2^{\frac 1 3}x^{-\frac 2 3}(1+2x)^{\frac 1 2}\\
&a_1=2^{\frac 4 3}x^{-\frac 2 3} (1+4x) \;,\qquad \Lambda = -2^{\frac 1 3}x^{-\frac 2 3}(1+4x)\;.
\end{align}
When the two charges are the same, there is a $\mathbb{Z}_2$ symmetry exchanging the two charges.
Since the gravity is insensitive to this exchange, only the combination $f^2+f^3$ couples to the metric component and $f^2-f^3$ decouples.
The decoupled mode is analyzed by considering the mass matrix in \eqref{massmatrix} with the eigenvector $(0,1,-1,0)$ for some $m^2$.
Due to the fact that $E_2=E_3$ and $X_2=X_3$, the only condition that we need to satisfy is
\begin{equation}
m^2-k^2+E_1 X_2^2 k=0\;.
\end{equation}
Therefore $m^2$ has the minimum value when $k=\frac {E_1 X_2^2} 2$, in which case $m^2=-\frac{E_1^2 X_2^4}{4}=-2^{\frac 4 3} x^{\frac 1 3}$.
The Breitenlohner-Freedman bound is $-\frac {a_1} 4 = -2^{-\frac 2 3}x^{-\frac 2 3} (1+4x)$.
Their ratio is $\frac{4x}{1+4x}$, which is always lower than 1.
That is, the mass squared is always above the bound.

Of course, it is possible that there are other modes that go below the bound.
But this turns out not to be the case.
To see this, let us evaluate the determinant \eqref{massmatrix} when the mass-squared $m^2$ takes the value $-\frac 1 4 a_1$, which is the Breitenlohner-Freedman bound.
Then this is a function of $k$ and $x$.
We can check that this function is always positive, meaning that the roots of the determinant equation, which are the possible values of the mass-squared, are all greater than the Breitenlohner-Freedman bound.

When all three charges are different, we have not been able to solve the equations analytically, so we resorted to a numerical method.
Given three charges, we first adjust the parameter
$\mu$ in \eqref{E:threechargemet} so that it gives an extremal black hole.
Then we evaluate the metric and the functions at the horizon and solve the mass matrix equation \eqref{massmatrix} for $m^2$.
In both $AdS_2\times S^3$ or $AdS_2\times \mathbb{R}^3$, however, no unstable modes are found for a large range of the three charges. The bound is always barely satisfied.

In this section, we studied stability of the three-charge black hole
in the near-horizon limit. 
As we saw in the previous section, the near-horizon analysis gives a 
sufficient but not necessary condition for the instability at $T=0$. 
However, the critical value of the Chern-Simons coupling is given 
correctly from the
near-horizon analysis. Thus, we expect that our conclusion in this section
would not be modified even if we perform the analysis in the full
black hole geometry.


\section*{Acknowledgments}

We thank Matthew Fisher, Koji Hashimoto, Gary Horowitz, 
Shunichiro Kinoshita, Alexei Kitaev, Masakiyo Kitazawa,
Hong Liu, Keiju Murata, Masaki Oshikawa,
Yuji Tachikawa and Tadashi Takayanagi for discussion.

H.~O. and C.-S.~P. are supported in part by DOE grant DE-FG03-92-ER40701.
H.~O. is also supported in part by the 
World Premier International Research Center Initiative of MEXT of Japan and
by a Grant-in-Aid for Scientific Research (C) 20540256 
of JSPS.
S.~N. is supported by the
Grant-in-Aid for the Global COE Program 
``The Next Generation of Physics, Spun from
Universality and Emergence" of MEXT of Japan.
S.~N. was also supported by the SRC Program of the KOSEF through the Center for Quantum Space-time of Sogang University with grant number R11-2005-021 and by the YST program of Asia Pacific Center for Theoretical Physics at the initial stage of the present work.

H.~O. thanks the Albert Einstein Institute in Golm, 
the Aspen Center for Physics,
the Galileo Galilei Institute in Florence, the Kavli Institute for
Theoretical Physics in Santa Barbara, and the Yukawa Institute
for Theoretical Physics in Kyoto for their hospitalities.


\begin{thebibliography}{99}
\parskip=-3pt

\bibitem{Hartnoll}
  S.~A.~Hartnoll,
  ``Lectures on holographic methods for condensed matter physics,''
  arXiv:0903.3246 [hep-th].

\bibitem{Herzog}
  C.~P.~Herzog,
  ``Lectures on holographic superfluidity and superconductivity,''
  J.\ Phys.\ A  {\bf 42}, 343001 (2009)
  [arXiv:0904.1975 [hep-th]].

\bibitem{Deserone}
  S.~Deser, R.~Jackiw and S.~Templeton,
  ``Three-dimensional massive gauge theories,''
  Phys.\ Rev.\ Lett.\  {\bf 48}, 975 (1982).

\bibitem{Desertwo}
  S.~Deser, R.~Jackiw and S.~Templeton,
  ``Topologically massive gauge theories,''
  Annals Phys.\  {\bf 140}, 372 (1982)
  [Erratum-ibid.\  {\bf 185}, 406.1988\ APNYA,281,409 (1988\ APNYA,281,409-449.2000)].




\bibitem{Chamblin:1999tk}
  A.~Chamblin, R.~Emparan, C.~V.~Johnson and R.~C.~Myers,
  ``Charged AdS black holes and catastrophic holography,''
  Phys.\ Rev.\  D {\bf 60}, 064018 (1999)
  [arXiv:hep-th/9902170].


\bibitem{Cvetic:1999ne}
  M.~Cvetic and S.~S.~Gubser,
  ``Phases of R-charged black holes, spinning branes and strongly coupled
  gauge theories,''
  JHEP {\bf 9904}, 024 (1999)
  [arXiv:hep-th/9902195].


\bibitem{Rey:2008zz}
  S.~J.~Rey,
  ``String theory on thin semiconductors: holographic realization of Fermi
  points and surfaces,''
  Prog.\ Theor.\ Phys.\ Suppl.\ {\bf 177}, 128 (2008).






\bibitem{Domokos}
  S.~K.~Domokos and J.~A.~Harvey,
  ``Baryon number-induced Chern-Simons couplings of vector and axial-vector
  mesons in holographic QCD,''
  Phys.\ Rev.\ Lett.\  {\bf 99}, 141602 (2007)
  [arXiv:0704.1604 [hep-ph]].




\bibitem{FF}
  P.~Fulde and R.~A.~Ferrell,
  ``Superconductivity in a strong spin-exchange Field,''
  Phys.\ Rev.\  {\bf 135}, A550 (1964).



\bibitem{LO}
  A.~I.~Larkin and Y.~N.~Ovchinnikov,
  ``Nonuniform state of superconductors,''
  Zh.\ Eksp.\ Teor.\ Fiz.\  {\bf 47}, 1136 (1964)
  [Sov.\ Phys.\ JETP {\bf 20}, 762 (1965)].




\bibitem{Alford}
  M.~G.~Alford, J.~A.~Bowers and K.~Rajagopal,
  ``Crystalline color superconductivity,''
  Phys.\ Rev.\  D {\bf 63}, 074016 (2001)
  [arXiv:hep-ph/0008208].

\bibitem{Rubakov}
  D.~V.~Deryagin, D.~Y.~Grigoriev and V.~A.~Rubakov,
  ``Standing wave ground state in high density, zero temperature QCD at large $N_c$,''
  Int.\ J.\ Mod.\ Phys.\  A {\bf 7}, 659 (1992).

\bibitem{Son}
  E.~Shuster and D.~T.~Son,
  ``On finite-density {QCD} at large $N_c$,''
  Nucl.\ Phys.\  B {\bf 573}, 434 (2000)
  [arXiv:hep-ph/9905448].



\bibitem{Brazovskii}
S.~A.~Brazovskii,
``Phase transition of an isotropic system to a nonuniform state,''
Sov.\ Phys.\ JETP, {\bf 41}, 85 (1975).

\bibitem{Hohenberg}
See, for example,
P.~C.~Hohenberg and J.~B.~Swift,
``Metastability in fluctuation driven first-order transitions: nucleation of lamellar phases,''
Phys. Rev. E {\bf 52}, 1828 (1995)
[arXiv:patt-sol/9501004],
 and the references therein.

\bibitem{Witten}
  E.~Witten,
  ``Anti-de Sitter space and holography,''
  Adv.\ Theor.\ Math.\ Phys.\  {\bf 2}, 253 (1998)
  [arXiv:hep-th/9802150].

\bibitem{Tachikawa}
  S.~Benvenuti, L.~A.~Pando Zayas and Y.~Tachikawa,
  ``Triangle anomalies from Einstein manifolds,''
  Adv.\ Theor.\ Math.\ Phys.\  {\bf 10}, 395 (2006)
  [arXiv:hep-th/0601054].


\bibitem{SwampOne}
  C.~Vafa,
  ``The string landscape and the swampland,''
  arXiv:hep-th/0509212.


\bibitem{SwampTwo}
  H.~Ooguri and C.~Vafa,
  ``On the geometry of the string landscape and the swampland,''
  Nucl.\ Phys.\  B {\bf 766}, 21 (2007)
  [arXiv:hep-th/0605264].

\bibitem{Klimov:1981ka}
  V.~V.~Klimov,
  ``Spectrum of elementary fermi excitations in quark gluon plasma. (In
  Russian),''
  Sov.\ J.\ Nucl.\ Phys.\  {\bf 33}, 934 (1981)
  [Yad.\ Fiz.\  {\bf 33}, 1734 (1981)].

\bibitem{Weldon:1982bn}
  H.~A.~Weldon,
  ``Effective fermion masses of order $gT$ in high temperature 
  gauge theories with exact chiral invariance,''
  Phys.\ Rev.\  D {\bf 26}, 2789 (1982).

\bibitem{Braaten:1990wp}
  E.~Braaten, R.~D.~Pisarski and T.~C.~Yuan,
  ``Production of soft dileptons in the quark-gluon plasma,''
  Phys.\ Rev.\ Lett.\  {\bf 64}, 2242 (1990).

\bibitem{LeBellac}
For review, see for example, M. Le Bellac, ``Thermal Field Theory,'' 
Cambridge University Press, Cambridge (2000).


\bibitem{gmt}
  J.~P.~Gauntlett, R.~C.~Myers and P.~K.~Townsend,
  ``Black holes of D = 5 supergravity,''
  Class.\ Quant.\ Grav.\  {\bf 16}, 1 (1999)
  [arXiv:hep-th/9810204].

\bibitem{kunz}
  J.~Kunz and F.~Navarro-Lerida,
  ``D = 5 Einstein-Maxwell-Chern-Simons black holes,''
  Phys.\ Rev.\ Lett.\  {\bf 96}, 081101 (2006)
  [arXiv:hep-th/0510250].



\bibitem{Erdmenger}
  J.~Erdmenger, M.~Haack, M.~Kaminski and A.~Yarom,
  ``Fluid dynamics of R-charged black holes,''
  JHEP {\bf 0901}, 055 (2009)
  [arXiv:0809.2488 [hep-th]].

\bibitem{Dutta}
  N.~Banerjee, J.~Bhattacharya, S.~Bhattacharyya, S.~Dutta, R.~Loganayagam and P.~Surowka,
  ``Hydrodynamics from charged black branes,''
  arXiv:0809.2596 [hep-th].

\bibitem{Torabian}
  M.~Torabian and H.~U.~Yee,
  ``Holographic nonlinear hydrodynamics from AdS/CFT with multiple/non-Abelian
  symmetries,''
  JHEP {\bf 0908}, 020 (2009)
  [arXiv:0903.4894 [hep-th]].

\bibitem{DSon}
  D.~T.~Son and P.~Surowka,
  ``Hydrodynamics with triangle anomalies,''
  arXiv:0906.5044 [hep-th].

\bibitem{Matsuo}
  Y.~Matsuo, S.~J.~Sin, S.~Takeuchi and T.~Tsukioka,
  ``Chern-Simons term in holographic hydrodynamics of charged AdS black hole,''
  arXiv:0910.3722 [hep-th].

\bibitem{Sahoo}
B.~Sahoo and H.-Y.~Yee, 
``Holographic chiral shear waves from anomaly''
arXiv:0910.5915 [hep-th]. 



\bibitem{Gunaydin}
  M.~Gunaydin, G.~Sierra and P.~K.~Townsend,
  ``The geometry of $N=2$ Maxwell-Einstein supergravity and Jordan algebras,''
  Nucl.\ Phys.\  B {\bf 242} (1984) 244.


\bibitem{fujii}
  A.~Fujii and R.~Kemmoku,
  ``D = 5 simple supergravity on $AdS_2 \times S^3$,''
  Phys.\ Lett.\  B {\bf 459}, 137 (1999)
  [arXiv:hep-th/9903231].




\bibitem{Gibbons:2002pq}
  G.~Gibbons and S.~A.~Hartnoll,
  ``A gravitational instability in higher dimensions,''
  Phys.\ Rev.\  D {\bf 66}, 064024 (2002)
  [arXiv:hep-th/0206202].


\bibitem{GubserOne}
  S.~S.~Gubser,
  ``Colorful horizons with charge in anti-de Sitter space,''
  Phys.\ Rev.\ Lett.\  {\bf 101}, 191601 (2008)
  [arXiv:0803.3483 [hep-th]].

\bibitem{GubserTwo}
  S.~S.~Gubser and S.~S.~Pufu,
  ``The gravity dual of a p-wave superconductor,''
  JHEP {\bf 0811}, 033 (2008)
  [arXiv:0805.2960 [hep-th]].


\bibitem{Cvetic}
  M.~Cvetic {\it et al.},
  ``Embedding AdS black holes in ten and eleven dimensions,''
  Nucl.\ Phys.\  B {\bf 558}, 96 (1999)
  [arXiv:hep-th/9903214].

\bibitem{Behrndt}
  K.~Behrndt, M.~Cvetic and W.~A.~Sabra,
  ``Non-extreme black holes of five dimensional $N = 2$ AdS supergravity,''
  Nucl.\ Phys.\  B {\bf 553}, 317 (1999)
  [arXiv:hep-th/9810227].









\end{thebibliography}
\end{document}